\documentstyle[prl,multicol,aps]{revtex}
\begin{document}
\input{epsf.tex}
\epsfverbosetrue

\title{Stability of Multi-hump Optical Solitons}

\author{Elena A. Ostrovskaya$^{1}$,  Yuri S. Kivshar$^{1}$,
Dmitry V. Skryabin$^{2}$, and William J. Firth$^{2}$}

\address{$^{1}$ Optical Sciences Center, The Australian National
University, Canberra ACT 0200, Australia\\
$^{2}$  Department of Physics and Applied Physics,
University of Strathclyde, Glasgow, G4 0NG, Scotland}
\maketitle

\begin{abstract}
We demonstrate that, in contrast with what was previously believed,
multi-hump solitary waves {\em can be stable}. By means of linear stability
analysis and numerical simulations, we investigate the stability of {\em two- and
three-hump solitary waves} governed by incoherent beam interaction in a
saturable medium, providing a theoretical background for the experimental
results reported by M. Mitchell, M. Segev, and D. Christodoulides
[Phys. Rev. Lett. {\bf 80}, 4657 (1998)].
\end{abstract}

\begin{multicols}{2}
\narrowtext

Self-guided optical beams, or {\em spatial optical solitons}, are
the building blocks of all-optical switching devices where  light itself
guides and steers light without fabricated waveguides \cite{phys_today}.
In the simplest case, a spatial soliton is  created by one beam of a
certain polarization and frequency, and it can  be viewed as a self-trapped
mode of an effective waveguide it induces in a  medium \cite{chiao}. When a
spatial soliton is composed of  two (or  more) modes of the induced
waveguide \cite{snyder_dyn}, its  structure becomes rather complicated, and
the soliton  intensity profile may  display several peaks. Such solitary
waves are  usually referred to as {\em  multi-hump solitons}; they have
been found for various nonlinear models of coupled fields \cite{multi_humped}.

In realistic (nonintegrable) physical models, solitary waves can become
unstable demonstrating self-focusing, decay, or a nonlinearity-driven
transition to a stable state, if the latter exists
\cite{Pelinovsky}. All these scenarios of soliton evolution  are initiated
by exponentially growing perturbations and they are attributed to  {\em
linear instability}.  It is usually believed that {\em all types} of
multi-hump  solitary waves {\em are linearly unstable}, except for the
special case of neutrally stable solitons in the integrable Manakov model
\cite{Kutusov}. On the contrary, recent experimental results
\cite{prl_multi} indicate the possibility of observing {\em stationary
structures} resembling multi-hump solitary waves. This naturally poses
a question:  {\em Were those observations only possible because of short
propagation distance and a small instability  growth rate?}  Definitely,
the experimental results challenge the conventional view on multi-hump
solitary waves in   different models of nonlinear physics.

The purpose of this Letter is twofold. First, we study the origin  of
multi-hump solitons supported by incoherent interaction of two optical
beams in a photorefractive medium. We find that multi-hump solitons
appear via bifurcations of one-component solitons and due to {\em the
process of hump multiplication}, when the intensity profile of a composite
soliton changes from single- to multi-humped with increasing power. Second, we perform numerical stability analysis of  two- and
three-hump solitary waves and also find analytically the instability
threshold for two-hump solitons. We reveal that two-hump solitary waves are
{\em linearly stable} in a wide region of their existence, whereas all
three-hump  solitons are {\em linearly unstable}, and that even linearly
stable multi-hump solitons may not survive collisions.

In the experiments \cite{prl_multi}, spatial multi-hump solitary waves
were generated  by incoherent interaction of two optical beams in a
biased photorefractive crystal. The corresponding model has been derived by
Christodoulides
{\em et al.} \cite{christ}, and it is described by a system of  two coupled
nonlinear
equations for the normalized beam envelopes, $u(x,z)$ and $w(x,z)$, which for
the purpose
of our current analysis can be written in the following form \cite{us}:
\begin{equation}
\label{eq_uw}
\begin{array}{l} {\displaystyle
i \frac{\partial  u}{\partial z}+\frac{1}{2}\frac{\partial^2  u}{\partial
x^2}+\frac{ u(| u|^2+| w|^2)}{1+s(| u|^2+| w|^2)}- u=0,
}\\*[9pt]
{\displaystyle
i \frac{\partial  w}{\partial z}+\frac{1}{2}\frac{\partial^2  w}{\partial
x^2}+\frac{ w(| u|^2+| w|^2)}{1+s(| u|^2+| w|^2)}-\lambda w=0,}
\end{array}
\end{equation}
where the transverse, $x$, and propagation, $z$, coordinates are measured
in the units of $(L_d/k)^{1/2}$ and $L_d$, respectively, $L_d$ is a
diffraction length, and  $k$ is the wavevector in the medium. The parameter $\lambda$ is a ratio of the nonlinear propagation constants, and $s$ is an
effective {\it
saturation parameter}.  For $s  \rightarrow 0$, the system (\ref{eq_uw})
reduces
to the integrable {\it Manakov  equations} \cite{Kutusov}.

We look for stationary, $z$-independent, solutions of Eqs.
(\ref{eq_uw}) with both components  $u(x)$ and  $w(x)$ real and vanishing as
$|x| \rightarrow \infty$. Different types of such two-component localized
solutions, existing for $0<\{\lambda,s\}<1$, can be characterized by the
total power, $ P(\lambda,s)=P_u+P_w$, where the partial powers,
$P_u=\int_{-\infty}^{\infty}|u|^2 dx$ and $P_w=\int_{-\infty}^{\infty}|w|^2
dx$, are integrals of motion. If
one of the components is small, i.e. $w/u \sim \varepsilon$,
Eqs. (\ref{eq_uw}) become decoupled and, in the leading order, the equation
for
the $u$-component  has a solution $u_0(x)$ in the form of a fundamental,
$sech$-like,  soliton with no nodes. The second equation can then be
considered as an
eigenvalue  problem for the ``modes'' $w_n(x)$ of a waveguide created by
the soliton  $u_0(x)$ with the  effective refractive  index profile $
u^2_0(x)/[1+s u^2_0(x)]$. Parameter $s$  determines the  total number of
guided modes and the cut-off value for each  mode,  $\lambda_n(s)$.
Therefore, a two-component {\em vector} soliton $(u_0,w_n)$ consists of a
fundamental soliton and an $n$th-order mode of the waveguide it induces in
the medium. Henceforward we denote such a composite  solitary wave by its
``state vector'':  $|0,n \rangle$.

\vspace{-2mm}
\begin{figure}
\setlength{\epsfxsize}{7.7cm}
\centerline{\epsfbox{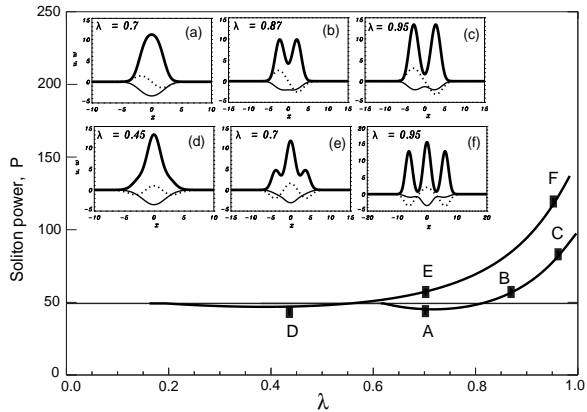}}
\caption{Soliton bifurcation diagram for $s=0.8$.  Horizontal line -
branch of the fundamental $u-$soliton. A-B-C - branch of  $| 0,1\rangle$ solitons. D-E-F
- branch of $| 0,2\rangle$ solitons. Inset: Transverse profiles of  $u$-
(thin), $w$- (dashed) fields,  and total intensity (thick), shown for
marked points.}
\label{fig1}
\end{figure}

On the $P(\lambda)$ diagram (for fixed $s$), continuous branches
representing $|0,n \rangle$ solitons  emerge at the  points of bifurcations
$\lambda_n(s)$ of
one-component solitons (see Fig. 1). It is noteworthy that the first-order mode
is in fact {\em the lowest possible mode of the waveguide} induced by the
fundamental soliton  $u_0(x)$.  This is because the state $| 0,0 \rangle$, node-less in both components, can exist only in the degenerate case
$\lambda=1$, when Eqs. (\ref{eq_uw}) have a family of equal-width solutions $u_0=A(x)\sin\theta$ and $w_0= A(x)\cos\theta$,
with arbitrary $\theta$, and amplitude  $A$ satisfying the scalar
equation,  $dA/dx=\pm s^{-1}[\log (1+s  A^2)-s(1-s)A^2]^{1/2}$.

Additionally, indefinitely many families of vector solitons $|m,n\rangle$,
where
$m\neq n \neq 0$,
can be formed as {\em bound states} of phase-locked $|0,n \rangle$ solitons
\cite{jeos,PhD}.
Although such states do contribute to the rich variety of the multi-hump
solitons existing in our model, we exclude them from our present consideration.

Families of vector solitons can be found by numerical relaxation
technique. Some results of our calculations are presented in Fig. 1, for
$|0,1 \rangle$ and $|0,2\rangle$  solitons found at $s=0.8$.
Observing the modification of soliton profiles with changing $\lambda$ (see
inset in Fig. 1),
one can see that the modal description of two-component solitons is valid only
near bifurcation points. For $\lambda \gg \lambda_n$, the amplitude
of an initially  small $w$-component {\em grows} and the soliton-induced
waveguide deforms.  {\it It is this purely nonlinear effect that gives rise to
the
existence of multi-hump  solitons.} In particular, two- and three-hump
solitons are members of the soliton families $|0,1 \rangle$ (branch
A-B-C) and  $|0,2\rangle$ (branch D-E-F) originating at different
bifurcation points. At $\lambda \sim \lambda_n(s)$, while
the $w$-component remains small, all $| 0,n\rangle$ solitons are {\it
single-humped}, as shown in Figs. 1(a,d). As the amplitude of $w$ grows with
increasing
$\lambda$, the total intensity profile, $I(x)= u^2_0(x)+w^2_n(x)$, develops
$(n+1)$
humps [see Figs. 1(b,e)], and at sufficiently large $\lambda$ the
$u$-component
itself becomes {\it multi-humped} [Figs. 1(c,f)]. The separation distance between
the soliton humps tends to infinity as $\lambda \rightarrow 1$.

To analyze the linear stability of multi-hump solitons, we  seek
solutions of  Eqs. (\ref{eq_uw})
in the form of weakly perturbed solitary waves:
$u(x,z)=u_0(x)+\varepsilon [F_u(x,z)+iG_u(x,z)]$
and $w(x,z)=w_n(x)+\varepsilon [F_w(x,z)+iG_w(x,z)]$, where
$\varepsilon \ll 1$. Setting  $F_{u,w}\sim f_{u,w}(x)e^{\beta z}$, $G_{u,w}\sim
g_{u,w}(x)e^{\beta z}$, one  can  obtain the following eigenvalue problem
(EVP)
\begin{eqnarray}
\hat{\mathcal{L}}_1 \hat{\mathcal{L}}_0\vec g=-\Lambda\vec
g,\label{evp1}\\
\hat{\mathcal{L}}_0 \hat{\mathcal{L}}_1\vec f=-\Lambda\vec
f. \label{evp2}\nonumber
\end{eqnarray}
Here $\vec g \equiv (g_u,g_w)^T$, $\vec f \equiv (f_u,f_w)^T$,
$\Lambda=\beta^2$,  and
\[
\hat{\mathcal{L}}_{0,1} = \left(\begin{array}{cc}
-\frac{1}{2}\frac{d^2}{d x^2} +1-a_{0,1} & b_{0,1}\\
b_{0,1} & -\frac{1}{2}\frac{d^2}{d x^2} +\lambda-c_{0,1}
\end{array}\right),
\]
where $a_0=c_0=I/(1+sI)$, $b_0=0$, $a_1=a_0+2u_0^2/(1+sI)^2$,
$c_1=c_0+2w_n^2/(1+sI)^2$, and $b_1=-2u_0w_n/(1+sI)^2$.

\vspace{-3mm}
\begin{figure}
\setlength{\epsfxsize}{6.0cm}
\centerline{\epsfbox{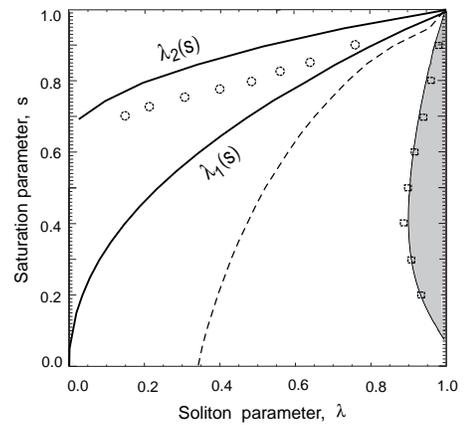}}
\vspace{3mm}
\caption{Existence and stability domains for two- and
 three-hump solitons. Shown are the existence thresholds $\lambda_1(s)$ and
$\lambda_2(s)$ for the $| 0,1\rangle$ and $| 0,2 \rangle$ soliton
families. Dashed - the line where $|0,1\rangle$ solitons become two-humped. Shaded - analytically obtained instability domain for two-hump solitons. Squares and circles- numerically
obtained
instability thresholds for $| 0,1\rangle$ and $| 0,2 \rangle$ solitons, respectively.}
\label{fig2}
\end{figure}

Because $\hat{\mathcal{L}}_{1}\hat{\mathcal{L}}_{0}$ and
$\hat{\mathcal{L}}_{0}\hat{\mathcal{L}}_{1}$
are adjoint operators with identical spectra, we can consider the
spectrum  of only one of
these operators, e.g. $\hat{\mathcal{L}}_{1}\hat{\mathcal{L}}_{0}$.
Considering the complex $\Lambda$-plane, it is straightforward to show that
$\Lambda\in (-\infty,-\lambda^2)$  is a continuum
part of the spectrum with unbounded eigenfunctions. Stable bounded
eigenmodes of
the
discrete spectrum (the so-called {\em soliton internal modes}
\cite{int_mode})  can have
eigenvalues
only inside the gap,  $-\lambda^2 < \Lambda <0$. The presence of
either positive or complex $\Lambda$ implies soliton
instability, because in this case there always exists at least one eigenvalue of the soliton spectrum with ${\rm Re} \beta >0$.

Numerical solution of the EVP (\ref{evp1}) shows that both $|0,1\rangle$ and $|0,2\rangle$ types of solitary wave solutions  {\it can be stable in a certain region of their existence domain}, see  Fig. 2. In the  case of $|0,1\rangle$ solitons, the appearance of the instability is related to  the fact
that close to the curve where the total intensity $I$ becomes two-humped
[dashed line in Fig. 2], a pair of internal modes split from the continuum
into the gap. As $\lambda$ grows, the corresponding, purely imaginary,
eigenvalues $\beta=\pm i
{\sqrt{|\Lambda(\lambda)|}}$ tend to zero, and at a certain
critical value $\lambda=\lambda_{\rm cr}(s)$, they coincide at $\beta=0$.
At this point, an eigenmode with positive eigenvalue $\Lambda$
emerges, thus generating linear
instability (see Fig. 3) with the instability growth rate $\beta = \sqrt{\Lambda(\lambda)}$. For $|0,2\rangle$ solutions, the dynamics of internal
modes can not be related in any obvious way with a change in the spatial
solitary  profiles,  nevertheless the scenario  of the instability
development is similar to that for two-hump solitons. The dependence
of $\beta$ on $\lambda$, for
$|0,1\rangle$ and $|0,2\rangle$ soliton families giving rise to  two-
and three-hump solitary waves, is shown  in Fig. 3 for $s=0.3$ and $s=0.8$,
respectively. A decline in the instability  growth rate as $\lambda\rightarrow
1$ (see Fig. 3) is caused by the fact that,  in this limit, all multi-hump
solitons decompose into a number of the neutrally  stable $|0,0\rangle$ solitons  separated by infinitely growing distance. Numerical analysis in the close vicinity of this limit is unfeasible due to lack of computational accuracy.

\vspace{-3mm}
\begin{figure}
\setlength{\epsfxsize}{6.5cm}
\centerline{\epsfbox{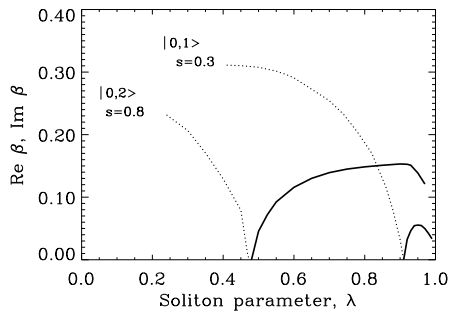}}
\vspace{3mm}
\caption{Instability eigenvalues vs. $\lambda$ for $| 0,1 \rangle$  and $|
0,2\rangle$ solitons; dashed line - ${\rm  Im} \beta$, bold line - ${\rm
Re}  \beta$.}
\label{fig3}
\end{figure}

Note that, within the gap of the continuous
spectrum, there exist several soliton internal modes not participating
in the development of the linear instability. Analysis of their origin
and influence on the soliton dynamics is beyond the scope of the
present Letter.

With the aid of analytical asymptotic technique \cite{Buryak},
it is possible to show that a perturbation mode with small but
positive  eigenvalue, and therefore the linear instability of a general localized
solution  $(u,w)$, appears if the functional $J(u,w),$ defined as
\begin{equation}\label{jacob_renorm}
J=\frac{P_u}{2s}\frac{\partial P_w}{\partial \lambda} -
\frac{P_w}{2s}\frac{\partial
P_u}{\partial \lambda} +
\frac{\partial P_u}{\partial s}\frac{\partial P_w}{\partial
\lambda}-\frac{\partial P_w}{\partial s}\frac{\partial P_u}{\partial
\lambda},\end{equation}
changes its sign.
The threshold condition $J=0$ is, in fact, the
 Vakhitov-Kolokolov stability criterion \cite{vk},
generalized for the case of two-parameter vector solitons.
In this case, it does not necessarily give a
threshold of leading instability \cite{makhankov}. Therefore, the
presence of other instabilities (which are not associated with the
condition $J=0$ and can have stronger growth rates) is still possible,
as in some other cases \cite{Dima}.

For two-hump solitons, we have been able to
locate the critical curve in $(\lambda,s)$-plane corresponding to the
condition  $J=0$. Superimposing this curve onto the numerically
calculated values $\lambda_{\rm cr}(s)$, we have found a remarkable agreement
between
the numerical and  analytical instability thresholds, as shown in Fig. 2.
 {\it This gives us the first example of the generalized Vakhitov-Kolokolov
criterion for the instability threshold of vector multi-hump solitary
waves.} For the whole family of $|0,2\rangle$ solutions, including three-hump solitons, it appears that $J\ne 0$
throughout the entire existence region. Thus,
appearance of instability of three-hump solutions is not associated with  the change of the sign of the functional $J$.

To analyze {\it long-term evolution} of multi-hump solitary waves, we perform
 numerical simulations  of the beam propagation for $|0,n\rangle$ solitons
within the existence domain $\lambda_n<\lambda<1$, at fixed $s$. First, we use
no perturbation so that the soliton instability can only develop from  
numerical noise. As long as the soliton  maintains its single-humped shape [see
corresponding profiles in Fig. 1(a,d)], it remains almost insensitive to
numerical noise.  Moreover, while the $|0,1\rangle$ solitons do become two-humped at
$\lambda<\lambda_{\rm cr}$, they still {\em remain stable in a wide domain
of their parameters}  until the linear instability  threshold is reached.
On the contrary, $|0,2\rangle$ solitons remain  single-humped up to the
 instability threshold value $\lambda=\lambda_{\rm cr}$, so that  {\em all  three-hump
solitons are indeed unstable}. Above the instability threshold (i.e. for
$\lambda _{\rm cr}< \lambda <1$), a two-hump soliton splits into two
independent single-humped beams
as a result of the instability developed from noise [see Fig. 4(a)],
whereas a three-hump soliton exhibits a more complex symmetry-breaking
instability, as shown in Fig. 4(b).

\vspace{-2mm}
\begin{figure}
\setlength{\epsfxsize}{7.5cm}
\centerline{\epsfbox{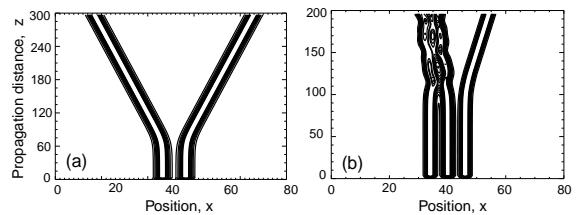}}
\vspace{3mm}
\caption{Noise-induced splitting of (a) two-hump soliton
[Fig.1(c)] and (b) three-hump soliton [Fig. 1(f)].}
\label{fig4}
\end{figure}

\vspace{-3mm}
\begin{figure}
\setlength{\epsfxsize}{7.5cm}
\centerline{\epsfbox{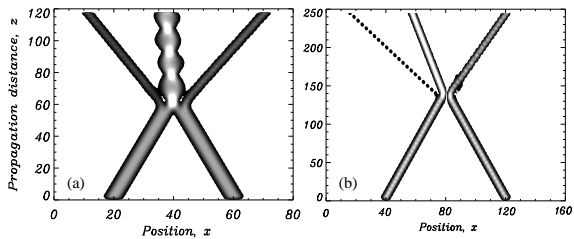}}
\vspace{3mm}
\caption{Collisions of (a) {\em linearly stable} $|0,1\rangle$ solitons
at $s=0.8$, $\lambda=0.72$ with the relative transverse velocity $v=0.05$,
and
(b) {\em linearly stable} $|0,2\rangle$
solitons at $s=0.8$, $\lambda=0.35$, with $v=0.09$.}
\label{fig6}
\end{figure}

Next, we propagate two-hump (at $s=0.3$) and three-hump (at $s=0.8$)
solitons perturbed by an eigenmode with the largest instability growth
rates, i.e. $\beta_{\rm max} \approx 0.055$ and $\beta_{\rm max} \approx
0.153$, respectively. We find that in the presence of $\sim 6 \%$ amplitude
perturbation, the diffraction-induced decay of a soliton can be stabilized by the nonlinearity, whereas its splitting is significantly
speeded up by the perturbation, compared with splitting due to
a numerical noise. 

To make a link between our stability analysis and experiment, we note that
for the experiment \cite{prl_multi} the diffraction length is defined as
$L_d = 2/sb$ and nonlinearity of the medium (SBN:60 crystal) is characterised by the parameter $b =
kr_{\rm eff} n_b^2 E_0$, where $r_{\rm eff}$ is the effective electro-optic
coefficient ($= 280$ pm/V), $n_b$ is the background refractive index $(=
2.3)$, and $E_0$ is the applied electric field ($\approx 2${\rm x}$10^5$ V/m).
For strong saturation we have $s \sim 1$ and $L_d \approx 0.2 $mm. Now, the
characteristic instability length $z_{\rm cr}$ can be defined through the
maximum growth rate $\beta_{\rm max}$ and, as a result, for two-hump solitons at $s=0.3$ we obtain $z_{\rm
cr} \approx 12.18$ mm. These estimates indicate that the instability, if it exists,
could be detected for two-hump solitons within the experimental
setup of Ref. \cite{prl_multi} and
therefore {\em stable two-hump solitons} have been indeed observed.

Importantly, three-hump solitons so far generated in the experiment belong to a {\em different class} of vector solitons which, in our notation, can be identified as $|1,2\rangle$ states. The extensive numerical analysis of soliton states $|1,2\rangle$ \cite{PhD} shows that all such solitons are linearly unstable. However, the observation of this instability is beyond the experimental parameters of Ref. \cite{prl_multi}. 

The complex structure of multi-hump solitons and nonintegrability of the
model (\ref{eq_uw}) result in a variety of collision scenarios, which are
{\em quite dissimilar} to the  collisions of multi-hump solitons of the exactly
integrable Manakov system \cite{Kutusov}. For instance, even linearly
stable vector solitons do not necessarily survive soliton collisions. In
Figs. 5(a,b) we show two examples  of  non-elastic interaction of {\it
linearly stable} $|0,1\rangle$ and  $|0,2\rangle$ solitons. 

In conclusion, we have analyzed, analytically and numerically, the stability
of multi-hump optical solitons in a saturable nonlinear medium. We have
found that multi-hump solitons are members of an extended class of vector solitons which can be {\em linearly stable} in a
wide region of their existence, although they may be destroyed in collisions.
We believe  that this is an important physical result that
calls for a revision of our understanding of the structure and stability
of many types of multi-hump solitary waves in nonintegrable
multi-component models, usually omitted in the analysis because of their
{\it a priori} assumed instability.

We thank  M. Segev, D. Christodoulides, M. Mitchell, and A.
Buryak   for useful  discussions. Yu.K. and E.O. are members of the
Australian Photonics
Cooperative  Research Centre.  D.S. acknowledges support from the Royal
Society
of Edinburgh  and British Petroleum.

\end{multicols}

\begin{references}

\bibitem{phys_today} M. Segev and G.I. Stegeman, Phys. Today {\bf 51}, 42 (1998).

\bibitem{chiao} R.Y. Chiao, E. Garmire, and C.H. Townes, Phys. Rev.
Lett. {\bf 13}, 479 (1964).

\bibitem{snyder_dyn} A.W. Snyder, S.J. Hewlett, and D.J. Mitchell, Phys.
 Rev. Lett. {\bf 72}, 1012 (1994).

\bibitem{multi_humped} See, e.g, I. A. Kol'chugina {\em et al.}, JETP Lett. {\bf 31}, 6 (1980); M. Haelterman and A.P. Sheppard, Phys.
Rev. E  {\bf 49}, 3376 (1994); J.M. Soto-Crespo {\em et al.}, Phys. Rev.
E {\bf  51}, 3547 (1995); A. Boardman {\em et al.}, Phys. Rev. A {\bf
52}, 4099  (1995); D. Michalache {\em et al.}, Opt. Eng. {\bf 35}, 1616
(1996); H.  He {\em et al.}, Phys. Rev. E {\bf 54}, 896 (1996).

\bibitem{Pelinovsky} See, e.g., D.E. Pelinovsky, V.V. Afanasjev, and Yu. S. Kivshar, Phys.
Rev. E {\bf 53},  1940 (1996).

\bibitem{Kutusov} S.V. Manakov, Sov. Phys. JETP {\bf 38}, 248 (1974); see
also V. Kutuzov {\em et. al.},  Phys. Rev. E {\bf 57}, 6056 (1998); N.
Akhmediev, W. Krolikowski, and A.W. Snyder, Phys. Rev. Lett. {\bf 81}, 4632
(1998).

\bibitem{prl_multi} M. Mitchell, M. Segev, and D.N. Christodoulides,
Phys. Rev. Lett. {\bf 80}, 4657 (1998).

\bibitem{christ} D.N. Christodoulides {\em et al.}, Appl. Phys. Lett.
{\bf 68}, 1763 (1996).

\bibitem{us} E.A. Ostrovskaya and Yu.S. Kivshar, Opt. Lett.
{\bf 23},  1268  (1998).

\bibitem{jeos} E.A. Ostrovskaya and Yu.S. Kivshar, J. Opt. B: Quantum
Semiclass. Opt. {\bf 1}, 77, (1999).

\bibitem{PhD} E.A. Ostrovskaya, PhD Thesis, The Australian National University, (1999).

\bibitem{int_mode} Yu.S. Kivshar {\em et al.}, Phys. Rev. Lett. {\bf 80}, 5032 (1998).

\bibitem{Buryak} A.V. Buryak, Yu.S. Kivshar, and S. Trillo, Phys. Rev.
Lett. {\bf 77}, 5210 (1996).

\bibitem{vk} M.G. Vakhitov and A.A. Kolokolov, Sov. Radiophys. {\bf 16},
783 (1973).

\bibitem{makhankov} V.G. Makhankov, Y.P. Rybakov, and V.I. Sanyuk,
Physics-Uspekhi {\bf 37}, 113 (1994).

\bibitem{Dima} D.V. Skryabin and W.J. Firth, Phys. Rev. E {\bf 58},
R1252 (1998);  P. Lundquist, D. R. Andersen, and Yu. S. Kivshar, Phys. Rev.
E  {\bf 57}, 3551 (1998); D. Mihalache, D. Mazilu, and L. Torner, Phys.
Rev. Lett. {\bf 81}, 4353 (1998).

\end{references}
\end{document}